%% file: color.tex
\documentclass{egpubl}
\usepackage{eg2020}
 
\ShortPresentation      
\usepackage[T1]{fontenc}
\usepackage{dfadobe}  

\usepackage{cite}  
\BibtexOrBiblatex
\electronicVersion
\PrintedOrElectronic
\ifpdf \usepackage[pdftex]{graphicx} \pdfcompresslevel=9
\else \usepackage[dvips]{graphicx} \fi

\usepackage{egweblnk} 

\title{Neural Smoke Stylization with Color Transfer
}

\author[F. Christen et al.]
{\parbox{\textwidth}{\centering
Fabienne Christen, Byungsoo Kim, Vinicius C. Azevedo and Barbara Solenthaler}
\\
{\parbox{\textwidth}{
\centering
ETH Z{\"u}rich
}}}

\include{include_cglsim}

\include{amsmath}
\usepackage[]{algorithm2e}
\input{math_commands.tex}

\begin{document}


\maketitle
\begin{abstract}
Artistically controlling fluid simulations requires a large amount of manual work by an artist. The recently presented transport-based neural style transfer approach simplifies workflows as it transfers the style of arbitrary input images onto 3D smoke simulations. However, the method only modifies the shape of the fluid but omits color information. In this work, we therefore extend the previous approach to obtain a complete pipeline for transferring shape and color information onto 2D and 3D smoke simulations with neural networks. Our results demonstrate that our method successfully transfers colored style features consistently in space and time to smoke data for different input textures.

\ccsdesc[500]{Computing methodologies~Physical simulation}
\ccsdesc[500]{Computing methodologies~Image processing}
\ccsdesc[500]{Computing methodologies~Neural networks}

\printccsdesc   
\end{abstract}

\input{introduction}
\input{preliminaries}
\input{method}
\input{results}
\input{conclusions}

\bibliographystyle{eg-alpha-doi}  
\bibliography{color}        


\end{document}

%% file: include_cglsim.tex
\usepackage{amsmath}

\usepackage{graphicx}
\usepackage[footnotesize,sl,SL,hang,tight]{subfigure}
\usepackage{multirow}
\usepackage{wrapfig}
\usepackage{xcolor}

\newcommand{\todo}[1]{}

\newcommand\hide[1]{}

%% file: math_commands.tex
\usepackage{amsmath,amsfonts,bm}

\def\eqref#1{equation~\ref{#1}}

\def\1{\bm{1}}

\DeclareMathAlphabet{\mathsfit}{\encodingdefault}{\sfdefault}{m}{sl}
\SetMathAlphabet{\mathsfit}{bold}{\encodingdefault}{\sfdefault}{bx}{n}

\DeclareMathOperator*{\argmin}{arg\,min}

%% file: introduction.tex
\section{Introduction}
\label{sec:Introduction}

Physically-based fluid simulations have become an integral part of special effects in movie production and graphics for computer games. However, artistic control of such simulations is not well supported and hence remains tedious, resource intensive and costly. 
Recent work on fluid control include target-driven optimization to find artificial forces to match given keyframes \cite{Treuille2003,Pan17} and velocity synthesis methods that allow augmentation with turbulent structures \cite{Kim08,Sato18}. With a neural flow stylization approach \cite{Kim2019}, more complex styles and semantic structures have been transferred in a post-processing step. Features from natural images are transferred onto smoke simulations, enabling general content-aware manipulations ranging from simple patterns to intricate motifs. The method is physically inspired, as it computes the density transport from a source input smoke to a desired target configuration. Stylizations from different camera viewpoints are merged to compute a 3D reconstruction of the smoke. While structural information is successfully transferred onto smoke data, color information was omitted. 
However, transferring texture information represents a valuable control tool for artists to change the appearance of a fluid.
Our work therefore extends the transport-based neural flow stylization of Kim et al. \shortcite{Kim2019} with a subsequent color optimization step that allows artists to control both style and color based on example images. The application is related to \cite{Jamriska15-SIG} that uses a flow-guided synthesis approach to transfer textures onto fluids.  

Flow stylization approaches extend existing image style transfer methods with spatio-temporal constraints. In the image processing literature, \cite{Gatys2016} automated the style transfer with neural networks and introduced several ways for the user to control the stylization effects \cite{Gatys2017}. Multiple follow-up works added new terms to the loss function of the original method to enhance the performance of the method. Histogram loss \cite{DBLP:journals/corr/WilmotRB17} prevents instabilities in the form of varying brightness and contrast throughout the stylized image and avoids washed out results, Laplacian loss \cite{Li:2017:LNS:3123266.3123425} preserves low-level details of the content image and a regularization term for photorealistic style transfer \cite{Luan2017} overcomes distortion problems that appear with the original loss function. \cite{Ruder2018} explored style transfer for video sequences ensuring the resulting frames to be temporally coherent and stable. 

The optimization of three-dimensional smoke data is possible through the use of a differentiable renderer. 
A differentiable renderer enables the computation of derivatives \cite{Loper:ECCV:2014}, and recent approaches presented a multipurpose differentiable ray tracer that integrates various parameters such as camera pose, scene geometry, materials, and lighting parameters \cite{Li:2018:DMC:3272127.3275109,Nimier19}. A lightweight and efficient differentiable renderer can be used in our case, as for flow stylization only the main flow structures need to be represented \cite{Kim2019}. 

%% file: preliminaries.tex
\section{Preliminaries}
\label{sec:Preliminaries}

Our approach for colorized smoke stylization is based on the original neural style transfer for images \cite{Gatys2016} and the transport-based neural style transfer for fluid simulations \cite{Kim2019}, which are briefly introduced in the following.

\subsection{Neural Style Transfer}
Neural Style Transfer (NST) is the process of synthesizing an image \(I\) from a style image \(I_S\) and a content image \(I_C\) through optimization using a convolutional neural network (CNN). The CNN is trained for natural image classification and its layers provide the feature space for the stylization. Using this CNN, the stylization can be formulated as an optimization problem as 
\begin{align}
I^* = \argmin_{I} \: \alpha \mathcal{L}_c(I, I_C) + \beta \mathcal{L}_s(I, I_S),
\label{eq:NSTloss}
\end{align}
where \(\mathcal{L}_c\) is the content loss, \(\mathcal{L}_s\) is the style loss and \(\alpha\) and \(\beta\) are weighing factors. 
The content loss is spatially aware and aims at preserving the overall structure of \(I_C\) in the synthesized image. The style loss on the other hand optimizes for style structures independently of their image position. Let \(F^{l}_I\) be the feature representation of image \(I\) on layer \(l\). The content loss \(\mathcal{L}_C\) and the style loss \(\mathcal{L}_S\) can then be formulated as
\begin{align}
\mathcal{L}_c(I, I_C) &= \sum_{l} (F^{l}_{I} - F^{l}_{C})^2 \label{eq:contentloss}\\
\mathcal{L}_s(I, I_S) &= \sum_{l} (G^l_I - G^l_{I_S})^2 \label{eq:styleloss}
\end{align}
where \(G^l_X = (F^{l}_{X})^T (F^{l}_{X})\) is the Gram matrix of the feature representation on layer \(l\) of an image \(X\).

\subsection{Transport-Based Neural Style Transfer} 

Transport-Based Neural Style Transfer (TNST) extends the original NST algorithm to transfer the appearance of a given image to flow-based smoke density. As opposed to NST where the stylized image is optimized, the optimization formulation for TNST outputs a velocity field. Consequently, no image pixels are modified directly. Instead, the input density \(d\) is transported by the optimized velocity field \(v^*\) to obtain the final stylized density \(d^*\). 
\(v^*\) and \(d^*\) are obtained through optimization analogously to Equation \ref{eq:NSTloss} with
\begin{align}
v^* &= \argmin_{v} \: \alpha \mathcal{L}_c(\mathcal{R}_\theta(\mathcal{T}(d, v)), I_C) + \beta \mathcal{L}_s(\mathcal{R}_\theta(\mathcal{T}(d, v)), I_S) \label{eq:TNSTloss}\\
d^* &= \mathcal{T}(d, v^*). \label{eq:finaldensity}
\end{align}
The transport function \(\mathcal{T}(d, v)\) advects the density by the given velocity. The renderer \(\mathcal{R}_{\theta} (d)\) renders a 2D greyscale image of the density \(d\) at viewpoint angle \(\theta\). Several viewpoints can be selected for the optimization to avoid distortions in the final stylized 3D density \(d^*\). The loss functions \(\mathcal{L}_C\) and \(\mathcal{L}_S\) maintain their definitions from Equation \ref{eq:contentloss} and \ref{eq:styleloss}. The content loss can be neglected in our case, as we only have a style image and there is no content that needs to be preserved.

To extend the single frame stylization to multiple frames in a time coherent way, TNST aligns the stylization velocities with the input velocities. This is done recursively for a pre-defined window size. 
Increasing the windows size enhances smoothness between consecutive frames, but simultaneously leads to larger memory requirements due to the recursive nature of the velocity alignment. 

%% file: method.tex
\section{Method}
\label{sec:Method}
\begin{figure}[h!]
\includegraphics[width=0.48\textwidth]{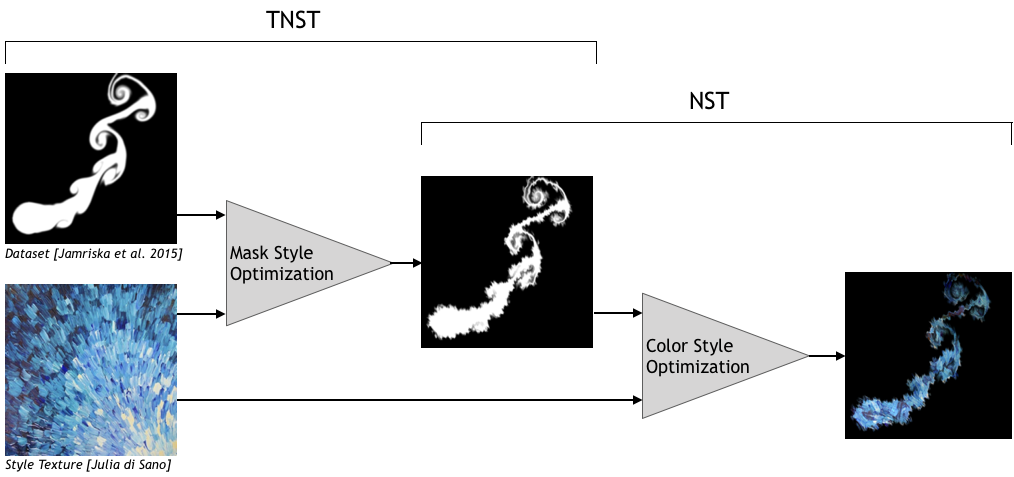}
\caption{Two step pipeline for colorized image style transfer.}\label{fig:pipeline}
\vspace{-12pt}
\end{figure}

Our method uses both NST and TNST as illustrated in Figure \ref{fig:pipeline}. In a first step, TNST is applied to the input frames of the smoke simulation to transfer structural information. This step corresponds to the approach of Kim et al. \shortcite{Kim2019}, and optimizes density values at each point. In a second step, we apply a color style optimization that modifies the color at each point while keeping the density values constant. 
%
%
%



\subsection{Color Style Optimization}
In the second step of the pipeline, color is added to the stylized mask \(d^*\) from the previous step. This part creates and optimizes color channels for \(d^*\), but does not further modify the density mask. The colorization process is performed using the original NST algorithm with a few alterations. Again, the desired style is given by the style image \(I_S\) and there is no content to preserve or transfer. Hence, we formulate the color style optimization as a simplified version of  Equation \ref{eq:NSTloss} without content loss:
\begin{align}
d^*_{RGB} = \argmin_{d} \: \mathcal{L}_s(\mathcal(\mathcal{R}_{RGB, \theta}(d), I_S).
\label{eq:color_optimization}
\end{align}
Since color information is now relevant for the optimization, the renderer \(\mathcal{R}_{RGB, \theta}(d)\) produces a 2D color image from viewpoint \(\theta\).

\begin{figure}[h!]
	\centering
	\subfigure[Output from density]{\label{fig:density_init}\includegraphics[width=0.155\textwidth]{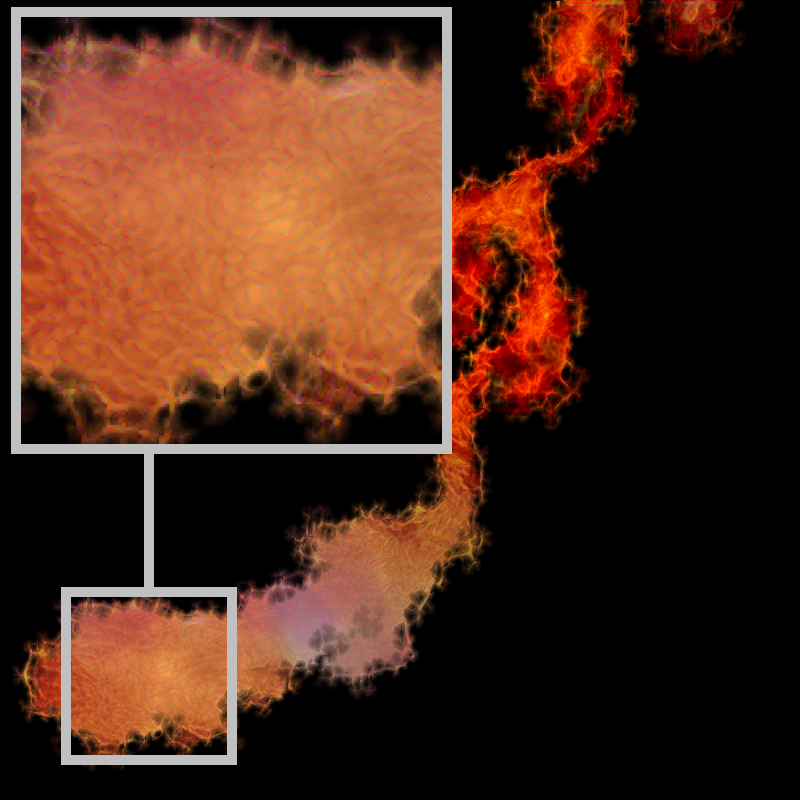}}
	\hfill
	\subfigure[Output from noise]{\label{fig:random_init}\includegraphics[width=0.155\textwidth]{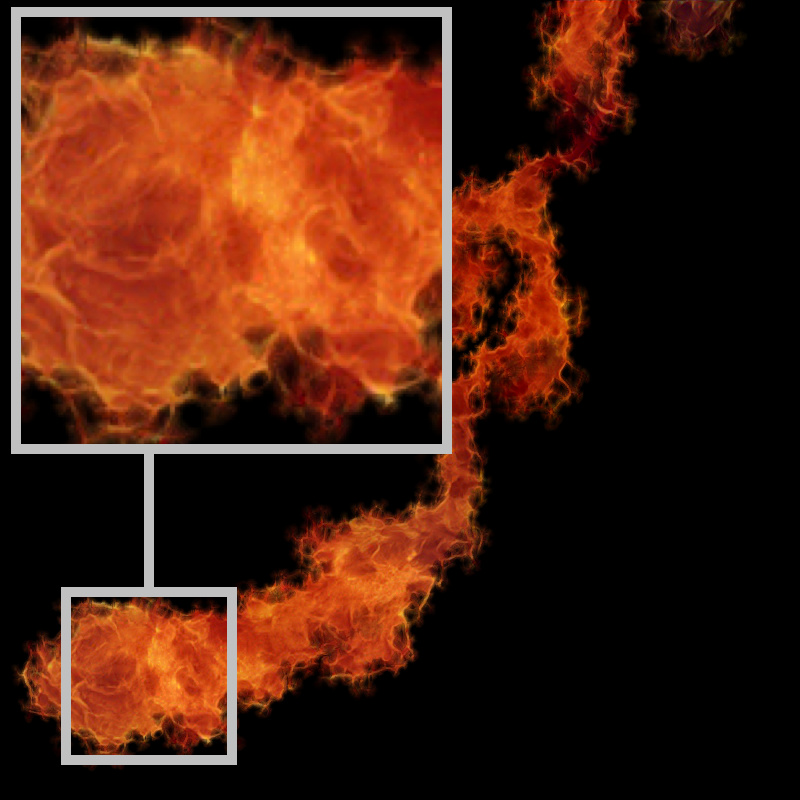}}
	\hfill
	\subfigure[Noise mask]{\label{fig:noise_mask}\includegraphics[width=0.155\textwidth]{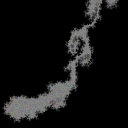}}
	\caption{Influence of initialization with a) stylized density only, b) density combined with noise as shown on c).}
	\label{fig:initialization}
	\vspace{-12pt}
\end{figure}

The proper initialization is crucial for the success of the color style optimization. As opposed to the original NST, there is no content loss, so any bias that is introduced in the initial condition can persist in the output. When starting the color style optimization from the stylized density \(d^*\), the initial pixel values of the area that will be stylized are close to white. This leads to washout effects as shown in Figure \ref{fig:density_init}. For the result in Figure \ref{fig:random_init} on the other hand, the stylized mask \(d^*\) is initially multiplied pointwise with white noise as shown on Figure \ref{fig:noise_mask}. This initial condition converges to a satisfying result. 

The color style optimization needs to be constrained to only optimize on the pixels that actually contain density. We obtain a guidance mask \(T^l\) by downsampling \(d^*\) to the size of each layer \(l\) that was selected for the style feature extraction and apply it to the style feature representation on layer \(l\) with \cite{Gatys2017}
\begin{align}
\hat{F}^l_{G(I)} = T^l \circ F^l_{G(I)},
\end{align}
where \(\circ\) denotes element-wise multiplication. This way of guiding the stylization will lead to some overflow at the boundaries, because the receptive fields of neurons near the boundaries can overlap the masked out regions. This overflow can be removed from the final stylized density by applying the guidance mask once in the end.

\subsection{Rendering}
\begin{figure}[h!]
	\centering
	\includegraphics[width=0.475\textwidth, trim=0 0 145 0, clip]{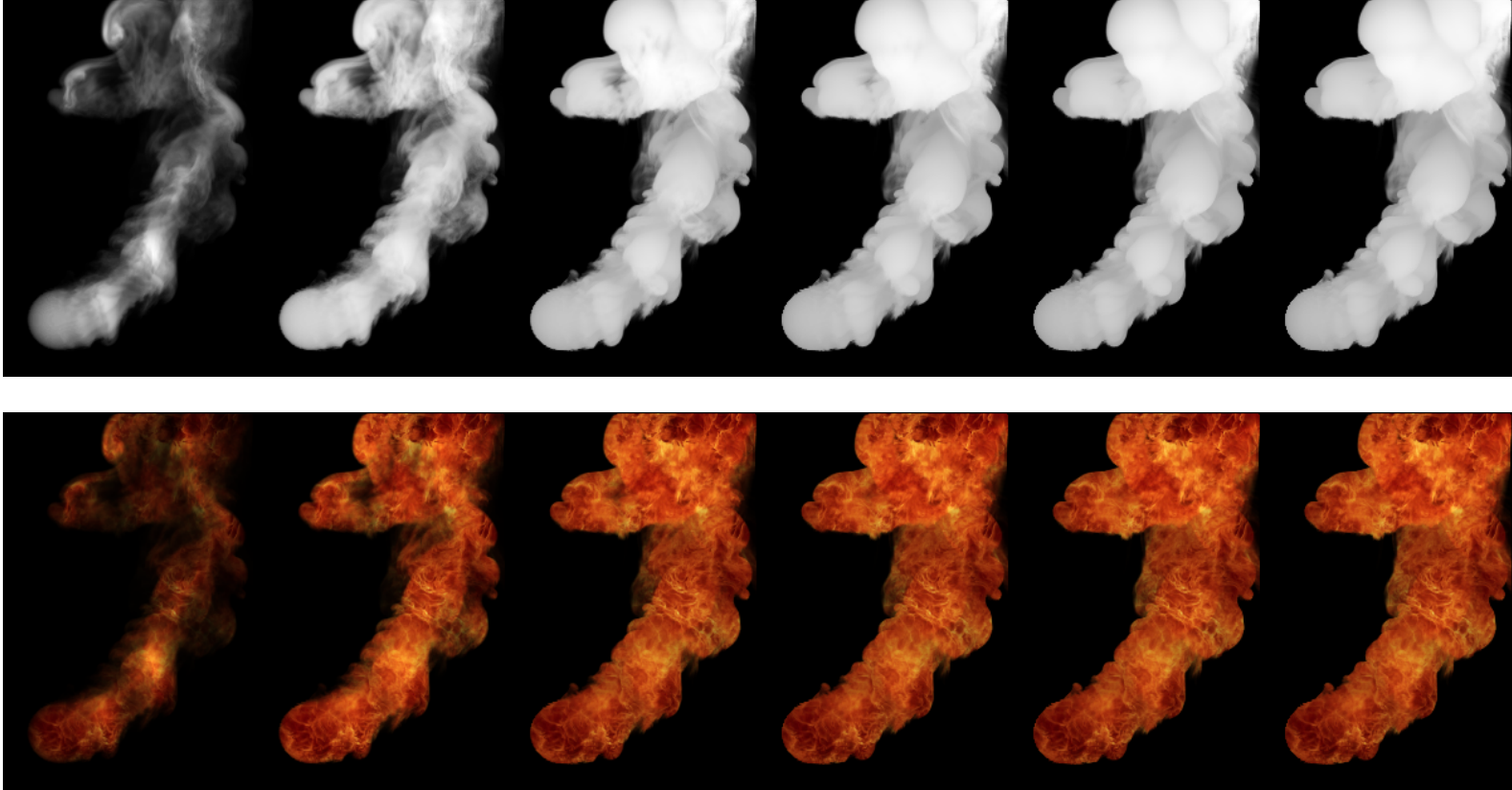}
	\caption{Output of the differentiable renderer without and with color channels for different transmittance factors $\gamma = [0.001, 0.1, 0.5, 1]$.}
	\label{fig:transmittance}
	\vspace{-12pt}
\end{figure}

Both renderers \(\mathcal{R}_\theta\) and  \(\mathcal{R}_{RGB,\theta}\) are part of the optimization pipeline and therefore need to be differentiable and lightweight. \(\mathcal{R}_\theta\) renders the smoke by calculating the pixel intensity along a ray in normal direction to the camera as proposed by \cite{Kim2019}. More specifically, the transmittance \(\tau(x)\) and the intensity \(I\) at each image pixel \(ij\) are defined \cite{Fong:2017:PVR:3084873.3084907} as
\begin{align}
\tau(x) &= e^{-\gamma \int_0^{r_{max}} d(r_{ij}) dr} \\
I_{ij} &= \int_0^{r_{max}} d(r_{ij}) \tau (r_{ij}) dr.
 \label{eq:transmittance}
\end{align}
The transmittance factor \(\gamma\) defines how much light is lost due to absorption and scattering, \(d(x)\) evaluates the amount of density at point \(x\), \(r_{ij}\) is the ray through pixel \(ij\) normal to the camera and \(r_{max}\) is the length of the ray. 
For the color style optimization, we extend this formulation to support color fields. The $C=\{ R,G,B \}$ emission values at each pixel \(ij\) are computed with
\begin{align}
C_{ij} &= \int_0^{r_{max}} C(r_{ij}) d(r_{ij}) \tau (r_{ij}) dr. 
\label{eq:color_intensity}
\end{align}
The density \(d_{ij}\) is multiplied into the emitted colors and can be seen as the emission factor at each point. 
Note that the \(RGB\) emission values are normalized to $[0..255]$.
The impact of the transmittance value onto the colorized result is illustrated in Figure \ref{fig:transmittance}.

\subsection{Controlling the Stylization}

\begin{figure}[h!]
	\centering
	\subfigure[Feature complexity increases with deeper layers of the CNN ('\textit{relu1\_1}'.. '\textit{relu4\_1}').]{\label{fig:relu}\includegraphics[width=0.4\textwidth, trim=0 0 180 0, clip]{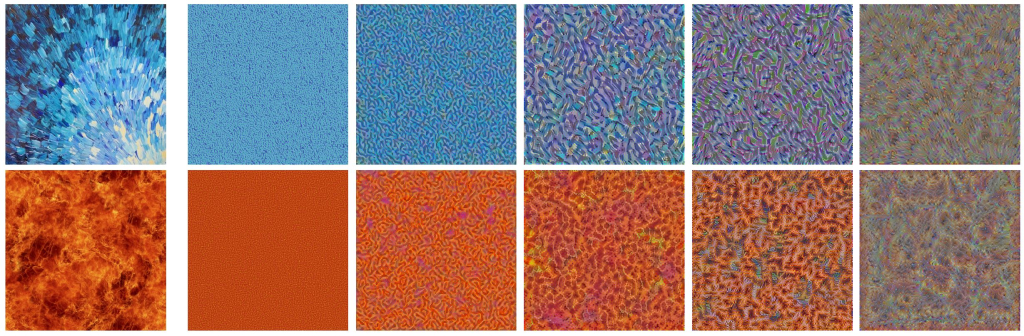}}
	\hfill
	\subfigure[Structure size can be controlled by tiling the input.]{\label{fig:tiles}\includegraphics[width=0.375\textwidth]{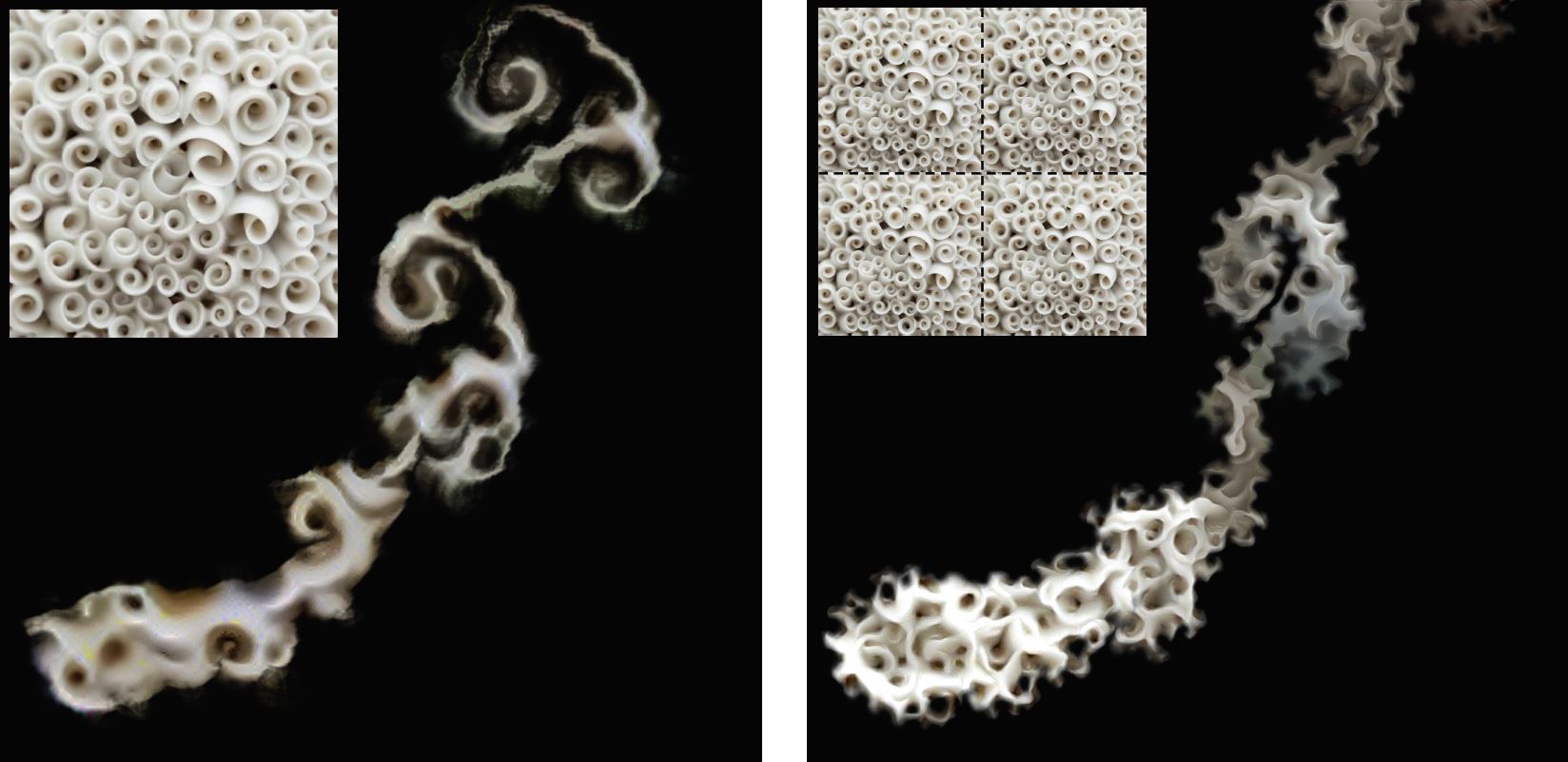}}
	\caption{Stylized structures can be controlled by selecting corresponding layers in the CNN and tiling the input image.}
	\label{fig:control}
	\vspace{-12pt}
\end{figure}

We used the VGG-19 network \cite{Simonyan15} for the feature extraction, which consists of 19 layers and has been trained for natural image classification.
%
The stylization can be controlled by selecting layers in the CNN. The deeper a layer is positioned in the CNN, the higher is the complexity of the extracted features, as illustrated in Figure \ref{fig:relu} for two different input images. The shallow layers optimize for low-level features, while deeper layers generate high-level features. 
The size of the stylized features depends on the size of the input image. Tiling can be used to progressively increase the input size to generate smaller scale structures as shown in Figure \ref{fig:tiles}.

%% file: results.tex
\section{Results}
\label{sec:Results}
%
%
%
\begin{figure}[h!]
\includegraphics[width=1.0\columnwidth
]{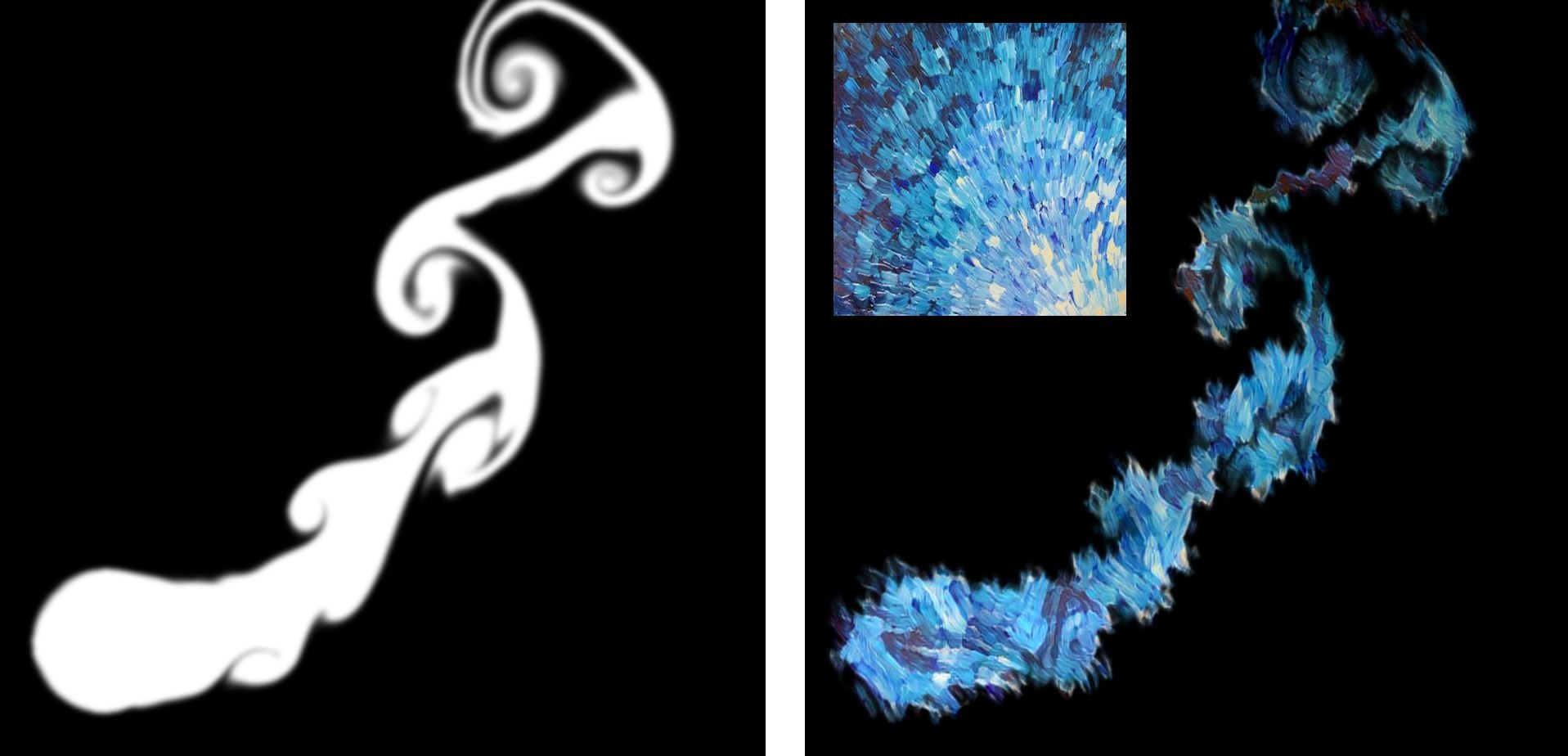}

\vspace{0.5mm}

\includegraphics[width=1.0\columnwidth
]{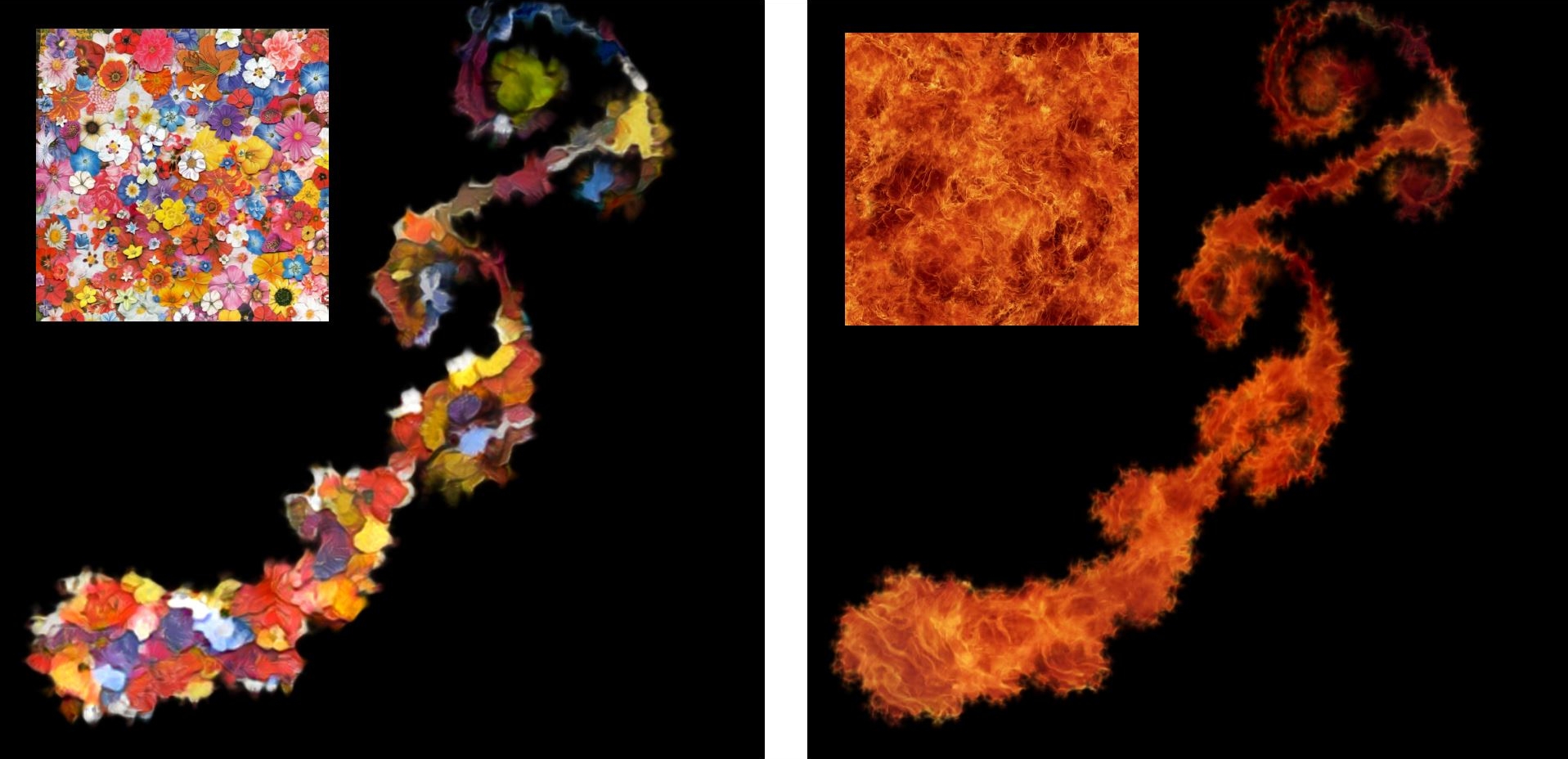}

\vspace{0.5mm}

\includegraphics[width=1.0\columnwidth
]{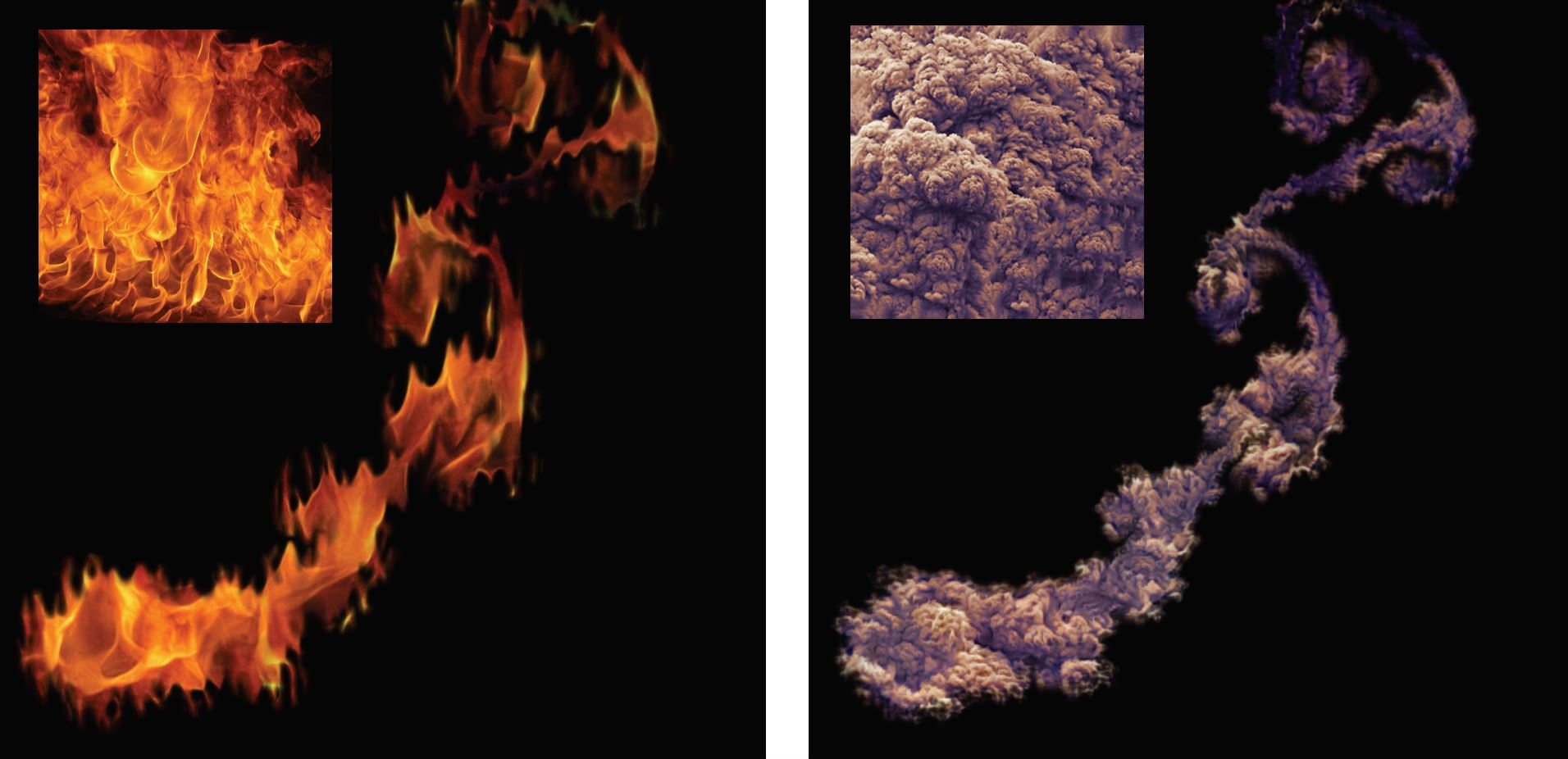}
  \caption{2D single frame color stylization applied to a data set of \cite{Jamriska15-SIG} using different input images (blue strokes, flower, flame, fire and volcano).}
  \label{fig:results}
  \vspace{-12pt}
\end{figure}
We implemented the stylization with TensorFlow and used the Adam optimizer with a learning rate of 0.5 and 1 for the 2D and 3D examples, respectively, for 300 iterations. For our results, we selected the layers '\textit{relu2\_1}' and '\textit{relu3\_1}' of the VGG-19 network for the feature extraction. 

We applied the style and color transfer to the 2D smoke data set of \cite{Jamriska15-SIG} using different input images as shown in Figure \ref{fig:results}. Color information is transferred coherently in space and time (see accompanying video sequences\footnote{\url{https://youtu.be/TyNlaBoP6oI}}), and hence complements the mask stylization of \cite{Kim2019}. 

\begin{figure}[h!]
\includegraphics[width=0.24\textwidth]{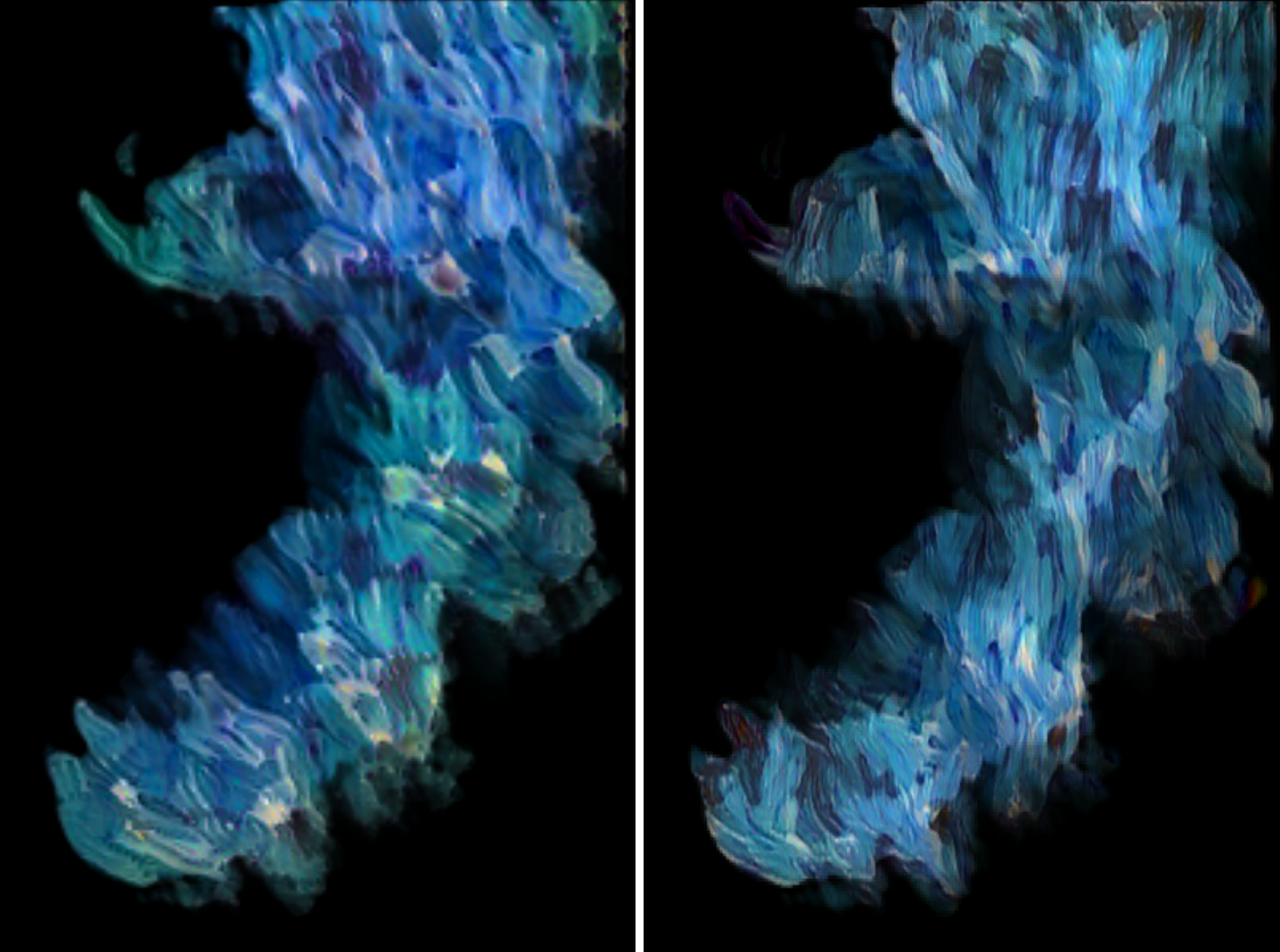}
\includegraphics[width=0.24\textwidth]{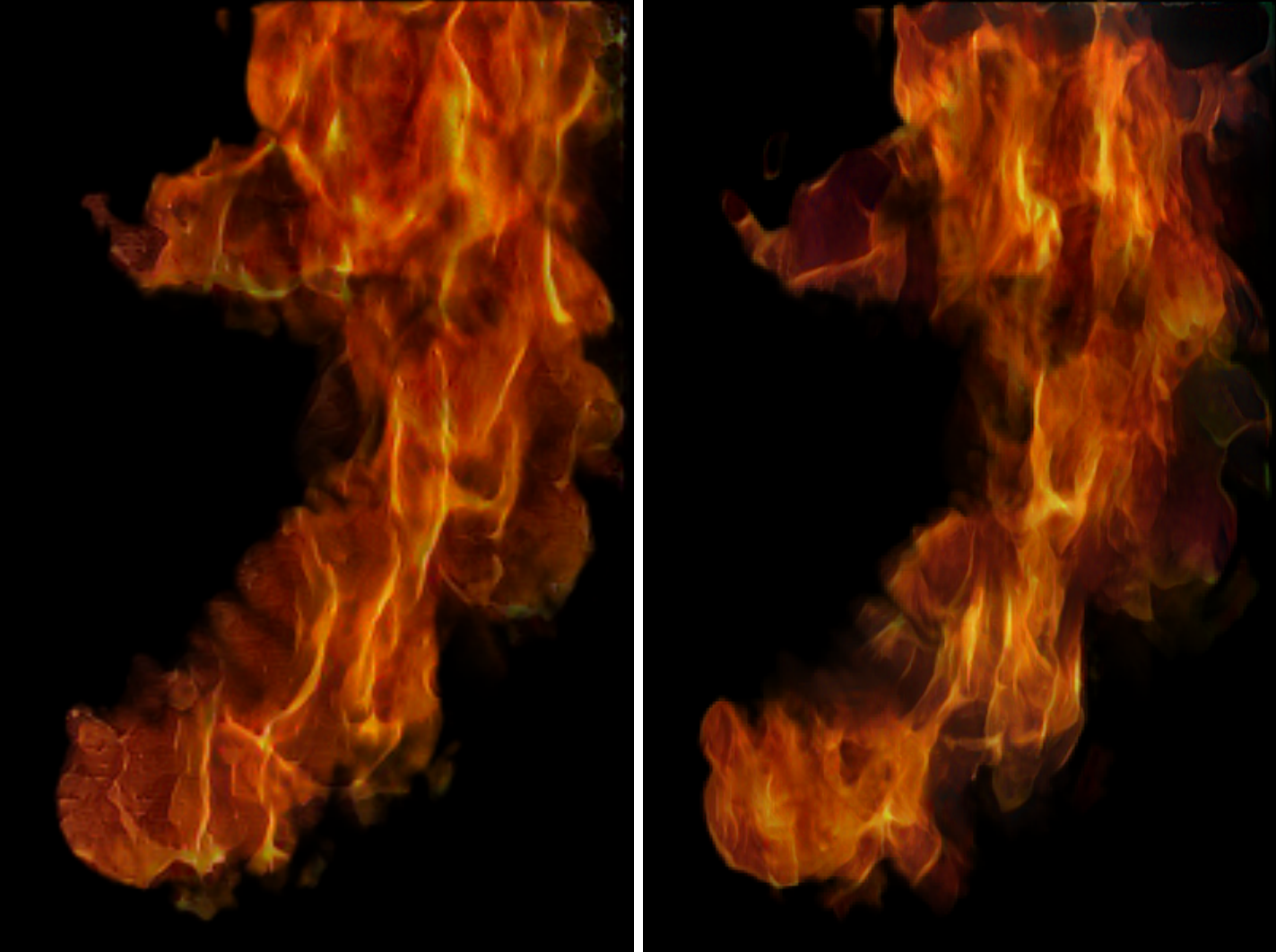}
  \caption{3D (left) and 2D (right) color stylization applied to a data set of \cite{Kim2019} for two input images (blue strokes and fire).}
  \label{fig:results3D}
  \vspace{-12pt}
\end{figure}

The 3D results were computed with a data set of \cite{Kim2019}, and shows the colorized outcome with the 3D pipeline that optimizes for multiple viewpoints as described in the original paper of \cite{Kim2019}. The lightweight and hence efficient differentiable color renderer is sufficient to capture the most relevant structures. We illustrate this by comparing the 3D results with their 2D counterparts in Figure \ref{fig:results3D}.

%% file: conclusions.tex
\section{Conclusion}
\label{sec:Conclusions}

In this work we extended an existing flow stylization approach by adding color transfer. The color stylization is coherent in space and time, and can be applied to 2D and 3D smoke densities. 
%
%
%
%
Our method directly optimizes for the stylized images during the training stage in an online fashion. Other research in the field of neural style transfer explores model-optimization based offline techniques. This type of style transfer technique moves the time intensive optimization into the phase of training the model, thereby gaining the advantage of stylizing images in a single forward pass. Using this optimization method would greatly reduce the time that the stylization takes.
Further, for the best outcome, the differentiable renderer that is used in the optimization should match the final high-quality rendering of the smoke. Our differentiable renderer could be adapted accordingly but at the cost of increased computation time.


\section{Acknowledgments}
\label{sec:Acknowledgments}
The authors would like to thank Ondrej Jamriska for sharing his dataset. This work was supported by the Swiss National Science Foundation (Grant No. 200021\_168997).